\documentclass[preprint,review, 11pt]{elsarticle}
\usepackage[margin=1in]{geometry}

\usepackage{graphicx}
\usepackage{float}
\usepackage{ctable}
\usepackage{algorithm}
\usepackage{wrapfig}
\usepackage{algpseudocode}
\usepackage{multirow}
\usepackage{amsfonts}
\usepackage{amsmath}
\usepackage{booktabs}
\usepackage{makecell}
\usepackage[ruled,vlined,linesnumbered,algo2e]{algorithm2e}
\usepackage{array}
\graphicspath{{figs/}}

\newcommand{\vect}[1]{\boldsymbol{#1}}
\graphicspath{{figs/}}

\journal{arXiv}

\graphicspath{{figs/}}

\begin{document}

\begin{frontmatter}

\title{Hierarchical Deep Learning with Generative Adversarial Network for Automatic Cardiac Diagnosis from ECG Signals}

\author{Zekai Wang}
\author{Stavros Stavrakis}
\author{Bing Yao\corref{cor1}}

\cortext[cor1]{Corresponding author: byao3@utk.edu; \\Zekai Wang and Bing Yao are with the Department of Industrial \& Systems Engineering, The University of Tennessee, Knoxville, TN, 37996 USA.\\Stavros Stavrakis is with University of Oklahoma Health Sciences Center, Oklahoma City, OK 73104 USA.}

\begin{abstract}
   	Cardiac disease is the leading cause of death in the US. Accurate heart disease detection is of critical importance for timely medical treatment to save patients' lives. Routine use of electrocardiogram (ECG) is the most common method for physicians to assess the electrical activities of the heart and detect possible abnormal cardiac conditions. Fully utilizing the ECG data for reliable heart disease detection depends on developing effective analytical models. In this paper, we propose a two-level hierarchical deep learning framework with Generative Adversarial Network (GAN) for automatic diagnosis of ECG signals. The first-level model is composed of a Memory-Augmented Deep auto-Encoder with GAN (MadeGAN), which aims to differentiate abnormal signals from normal ECGs for anomaly detection. The second-level learning aims at robust multi-class classification for different arrhythmias identification, which is achieved by integrating the transfer learning technique to transfer knowledge from the first-level learning with the multi-branching architecture to handle the data-lacking and imbalanced data issue. We evaluate the performance of the proposed framework using real-world medical data from the MIT-BIH arrhythmia database. Experimental results show that our proposed model outperforms existing methods that are commonly used in current practice.

\end{abstract}

\end{frontmatter}

\section{Introduction}
\label{sec1}
Heart disease is the leading cause of death in the US. It affects about 85.6 million people and leads to more than \$320 billion in annual medical costs \cite{virani2021heart}. It is of critical importance to develop accurate and reliable heart disease diagnoses for timely medical treatments to save patients' lives \cite{yao2021spatiotemporal,yao2016physics}. The heart rhythm is generated by the excitation, propagation, and coordination of electrical signals from the cardiac cells across different heart chambers. A normal cardiac cycle starts with the activation of the sinoatrial node, from where the cardiac electrodynamics spreads out through the atria. The electrical wave then arrives at the atrio-ventricular node and propagates through the bundle of His toward Purkinje fibers, leading to the electrical depolarization and repolarization of the ventricles to complete the cycle. The resulting electrical signals on the body surface are described by the electrocardiogram (ECG), which consists of a P-wave, QRS-complex, and T-wave \cite{hurst1998naming}. Changes in electrophysiological properties will vary the propagation pattern of electrodynamics and lead to different types of conduction abnormalities and/or cardiac arrhythmias manifested in the variation of ECG waveform patterns \cite{garabelli2016comparison,borleffs2009predicting}.

In recent years, rapid advancements in wearable sensing and information technology facilitate the effective monitoring of patients’ heart health conditions \cite{zhu2018optimal,liu2020gaining,yao2021constrained,yang2020network,iranzad2021gradient,yao2020spatiotemporal,xie2021physics,xie2022physics}. Routine use of ECG is the most common method for physicians in everyday clinical practice to assess the electrical activities of the heart and detect possible abnormal cardiac conditions. Physicians generally identify the cardiac arrhythmia by checking the ECG waveforms with naked eyes. This can be time-consuming and may require extensive human resources. Additionally, ECG misinterpretation may happen especially when there exists a large amount of data to inspect, leading to possible misdiagnosis of fatal heart disease \cite{schull2006risk}. Auto arrhythmia detection based on machine learning algorithms can provide important assistance to physicians \cite{rizwan2020review}. However, although ECG signals contain rich information associated with the electrophysiological condition of the heart, the research on fully utilizing ECGs for reliable data-driven disease detection poses several challenges including

(1) \textbf{Nonlinear and nonstationary dynamics}:  Real-world cardiovascular systems are featured with nonlinear and nonstationary dynamics from the complicated interactions of many interconnected parts such as ion channels and gap junctions to perform cardiac functions, generating ECG signals with nonlinear waveforms. Traditional statistical and machine learning methods depend heavily on manual feature engineering of such waveform data, which generally consists of two stages \cite{minchole2019machine}: human experts extract useful features from raw ECGs at the first stage and then employ machine learning algorithms on the handcrafted features to generate predictive results at the second stage. However, this procedure is restricted by the data quality and human expert knowledge \cite{guglin2006common}, and may result in information loss, which lacks the potential for real clinical implementation. Thus, new algorithms that are able to effectively and automatically extract useful features are urgently needed for reliable heart disease identification.

(2) \textbf{Lack of training labels and imbalanced data issue}:  Most existing data-driven models for ECG analysis are achieved through supervised learning, which requires a large volume of annotated ECG cycles (with diagnostic labels such as normal, abnormal, or specific types of arrhythmia). However, the annotation process requires cardiologists to manually inspect the ECG signals and assign a label to each different pattern, which is time-consuming and labor-intensive. Additionally, it is impractical to collect enough data for each type of disease-altered signals in order to meet the requirement for sufficient supervised training. This is due to the fact that data associated with abnormal heart conditions is significantly less than data from healthy people. Moreover, the occurrence rate of different arrhythmias is highly diverse. Data-driven predictive modeling based on such imbalanced data tends to ignore the minority classes, leading to unsatisfactory detection performance.  As such, new methods that can effectively model the ECGs and account for the data-lacking and imbalanced data issues are needed for reliable disease identification.

This paper proposes a hierarchical deep learning framework with Generative Adversarial Network (GAN) to investigate ECG signals for automatic identification of different types of arrhythmias. We first propose a Memory-Augmented Deep auto-Encoder with Generative Adversarial Network (MadeGAN) to achieve the first-level anomaly detection (i.e., binary classification for normal and abnormal signals). Second, we employ the transfer learning technique to transfer knowledge learned from the first-level training for second-level multi-class classification to identify different types of arrhythmias. In addition, in the second-level network, we adapt the multi-branching architecture developed in our prior work \cite{wang2021multi} to solve the imbalanced data issue among different types of heart diseases. We evaluate our proposed hierarchical deep learning framework using the data from the MIT-BIH arrhythmia database \cite{moody2001impact}. Experimental results show that our proposed method significantly outperforms existing approaches that are commonly used in current practice. 


\section{Research Background} 
\label{sec2}
A variety of statistical and machine learning algorithms have been developed for ECG data analysis and pattern recognition \cite{wang2013enabling}. For example, Yang \textit{et al} \cite{yang2013spatiotemporal} developed a dynamic spatiotemporal warping algorithm to measure dissimilarities between ECG signals and further employed the spatial embedding to transform the warping dissimilarity matrix into feature vectors for myocardial infarction classification. Bertsimas \textit{et al} \cite{bertsimas2021machine} utilized the XGBoost algorithm to capture disease-altered patterns in ECG cycles for heart disease prediction. Wavelet-based and recurrence analysis approaches have also been widely implemented to learn waveform features for ECG classification \cite{addison2005wavelet,yao2017characterizing,chen2019heterogeneous}. Lyon \textit{et al} \cite{lyon2018computational} investigated the linear and quadratic discriminants, support vector machine, random forest, and Bayesian network for heartbeat classification from ECG signals. A comprehensive review of statistical and machine learning methods in ECG detection can be found in \cite{minchole2019machine}.  However, most existing traditional data-driven methods depend heavily on manual feature engineering, which is a labor-intensive trial-and-error process and is generally limited by human expert knowledge \cite{guglin2006common, schlapfer2017computer}.

Deep Neural Network (DNN) is another powerful tool that has achieved promising results in the area of data-driven disease detection \cite{lecun2015deep}. Unlike conventional statistical and machine learning methods, the main advantage of DNNs is that they do not require explicit feature engineering. Instead, feature extraction is automatically achieved by intermediate layers of the network. It has been demonstrated that DNN-based features are more informative than handcrafted features for arrhythmia detection \cite{hong2019combining,galloway20195105}. As such, a variety of DNN models including convolutional neural networks (CNNs) \cite{albawi2017understanding} have been designed for arrhythmia detection and have outperformed conventional statistical methods \cite{ebrahimi2020review,murat2020application}. For example, Hannun \textit{et al} \cite{hannun2019cardiologist} employed 1D CNN to classify 12 rhythm classes and achieved high performance that is comparable to the diagnostic results provided by cardiology experts. Li \textit{et al} \cite{li2018combining} combined a 2D CNN and a distance matrix to classify congestive heart failure. Shashikumar \textit{et al} \cite{shashikumar2018detection} developed an attention-based model with a 2d CNN as the feature extractor and a bidirectional recurrent neural network to capture the temporal pattern in ECG signals.

However, most existing deep learning algorithms for ECG analysis are based on supervised learning, which requires a large volume of annotated ECG signals and also suffers from the problem of extremely imbalanced data. Thus, the application of unsupervised and semi-supervised learning in ECG analysis has been increasingly investigated. For example, Auto-Encoder (AE), a semi-supervised deep learning technique, has been widely used to study ECG data by extracting critical low-dimension representation of the raw signals for disease prediction \cite{xia2018automatic, xiong2016stacked}. Furthermore, GAN-based framework, another semi-supervised learning technique to capture inherent data distributions \cite{goodfellow2014generative,li2021augmented}, has been applied in ECG analysis. For example, Zhou \textit{et al} \cite{zhou2019beatgan} developed a BeatGAN structure to model ECG signals for anomaly detection. Wang \textit{et al} \cite{wang2019ecg} employed an auxiliary classifier GAN for data augmentation to handle the imbalanced issue. Shin \textit{et al} \cite{shin2020decision} integrated the AnoGAN framework \cite{schlegl2017unsupervised} with a decision boundary-based model for ECG anomaly detection. However, most existing semi-supervised deep learning methods mainly focus on differentiating the abnormal ECGs from normal ones (i.e., binary classification) and they are not able to perform multi-class classification to identify different types of cardiac arrhythmia. Thus, novel analytical models are urgently needed to efficiently handle the imbalanced data issue and the data lacking problem for both robust anomaly detection and accurate disease identification from ECG signals.

\begin{figure*}
	\begin{center}
		\includegraphics[width=5.5in]{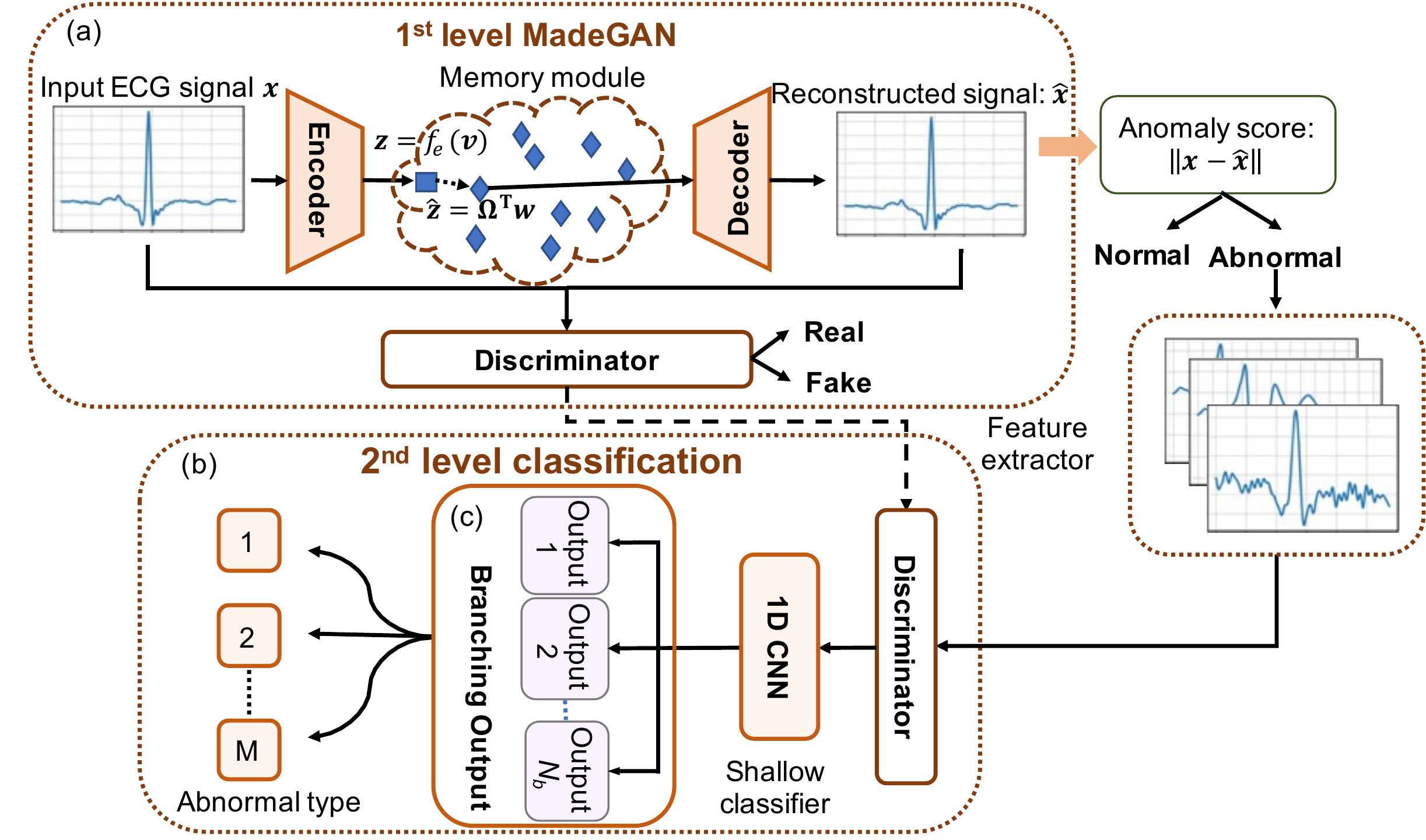}
		\caption{{The proposed two-level hierarchical deep learning framework: (a) first-level MadeGAN for anomaly detection; (b) second-level classification for arrhythmia type identification; (c) Multi-branching output.}}
		\label{Fig:2levelAD}
	\end{center}      	
\end{figure*}

\section{Research Methodology} 
\label{sec3} 
As shown in Fig. \ref{Fig:2levelAD}, this section presents the proposed hierarchical deep learning framework for automatic ECG diagnosis. We denote a single ECG cycle as $\vect{x} \in \mathbb{R}^{d_x \times 1}$, where $d_x$ denotes the dimensionality of $\vect{x}$. Each ECG cycle is associated with a multiclass label $y$. As such, each training data point can be described by the tuple $( \vect{x} , y)$ with $y=0$ indicating normal signal and $y=1,2,\dots, M$ corresponding to other different types of arrhythmias. Our objective is to first differentiate abnormal ECG signals from normal ones (i.e., first-level anomaly detection) and then classify the abnormal signals into different types of arrhythmias (i.e., second-level multi-class classification). Specifically, we propose a Memory-Augmented Deep auto-Encoder with Generative Adversarial Network (MadeGAN) to achieve the first-level anomaly detection. The second-level classification network is constructed by integrating a shallow classifier with the part of trained discriminator from the first-level learning (i.e., transfer learning to handle the data-lacking problem) and a multi-branching layer (to handle the imbalanced data issue).

\subsection{MadeGAN for First-level Anomaly Detection}
The goal of first-level MadeGAN is to differentiate abnormal ECG cycles from normal ones. Specifically, MadeGAN is trained with normal samples and aims to generate similar reconstructed signals for normal ECG inputs with small anomaly scores and dissimilar reconstructions for abnormal ECGs with bigger anomaly scores. As shown in Fig. \ref{Fig:2levelAD}(a), the MadeGAN is designed by integrating semi-supervised learning based on memory-augmented deep auto-encoder (MemAE) \cite{gong2019memorizing,park2020learning} with adversarial training, which will be detailed in the following subsections.

\subsubsection{Semi-Supervised Learning based on Memory-Augmented Deep Auto-Encoder (MemAE)}

The MemAE consists of three key modules: an encoder to encode the input signal $\vect{x} \in\mathcal{X}$ and generate a query latent vector $\vect{z} \in \mathcal{Z}$, a decoder to reconstruct signal $\hat{\vect{x}} $ from the latent space, and a memory module with a prototypical database to store the encoded patterns of normal ECGs. In other words, MemAE is composed by adding a memory module into an Auto-Encoder (AE).

An AE leverages feedforward neural networks for representation learning where the input is the same as the output. A traditional AE only contains an encoder and a decoder. The encoder is characterized by a function $f_e (\cdot)$ from original data domain $\mathcal{X}  \subset \mathbb{R}^{d_x \times 1}$ to a latent manifold domain $\mathcal{Z}\subset R^{d_z\times1}$, i.e., $f_e (\cdot): \mathcal{X} \rightarrow \mathcal{Z}$, which captures the critical information of the original signal $\vect{x}$ by designing a latent vector $\vect{z}$ with reduced dimensionality (i.e., $d_z<d_x$) through 
\begin{eqnarray}
	\vect{z}=f_e(\vect{x},\vect{\theta}_e)
\end{eqnarray}	 
where $\vect{\theta}_e$ denotes the parameter set of the encoder. The decoder is defined by the function $f_d (\cdot)$ to convert $\vect{z}$ back to the original space and generate the reconstructed signal as 
\begin{eqnarray}
	\hat{\vect{x}} =f_d ( \vect{z},\vect{\theta}_d)=f_d ( f_e(\vect{x},\vect{\theta}_e),\vect{\theta}_d)
\end{eqnarray}
where $\vect{\theta}_d$ denotes the parameter set of the decoder. The objective of AE is to effectively compress the input data $\vect{x} $ into a latent-space representation $\vect{z} $ and reconstruct a signal $\hat{\vect{x}} $ that is close to $\vect{x} $ from the encoded representation. This optimization objective is generally achieved by minimizing the following reconstruction error:
\begin{eqnarray}
	\label{AE}
	\mathcal{L}_{ed}(\vect{\theta}_e,\vect{\theta}_d)&=&\frac{1}{N}\sum_{i=1}^N\|\hat{\vect{x}}-\vect{x}\|^2
\end{eqnarray}
where $N$ denotes the size of the training set. If the AE is trained purely based on normal data, it is expected that bigger reconstruction errors will be produced for abnormal input signals than the normal inputs. As such, AE has been widely applied for anomaly detection in various areas \cite{gong2019memorizing,park2020learning,zhou2017anomaly, shi2022lstm}.

However, traditional AE-based approaches for anomaly detection often fail to explicitly consider variations in normal signals. In particular, nonlinear and nonstationary dynamics are inherent in the real-world cardiovascular system, generating ECG signals with nonlinear waveforms. There are variations in the shape, amplitude, and phase among different ECG cycles even under normal conditions. Due to the powerful representation capability of DNN, the AE trained by normal signals with diverse patterns tends to generalize well and may reconstruct abnormal signals with small errors \cite{gong2019memorizing,park2020learning}. To mitigate the drawback of traditional AEs, we leverage a memory module to recognize the diversity of normal patterns and reduce the generalizability of AE to abnormal signals as inspired by the work from \cite{gong2019memorizing,park2020learning}. 

The memory module is added in between the encoder and decoder as shown in Fig. \ref{Fig:2levelAD}(a) and is formed by a memory database with a matrix of $\vect{\Omega}\subset\mathbb{R}^{N_{\Omega}\times d_z}$ to store the representative latent features of normal ECG signals, where $N_{\Omega}$ denotes the number of prototypical vectors stored in the memory module. Given a latent variable $\mathbf{z}$, the memory module will generate a retrieved latent representation according to 
\begin{eqnarray}
	\label{zhat}
	\hat{\mathbf{z}} = \mathbf{\Omega}^T\vect{w}\ = \sum^{N_{\Omega}}_{i=1}{w_i\vect{\omega}_i}	
\end{eqnarray}
where $\vect{\omega}_i$ denotes the $i$th memory element in the memory database $\vect{\Omega}$, and $\vect{w}$ is a weight vector with $w_i\geq 0$ and $\sum_{i=1}^{N_{\Omega}}w_i=1$ to characterize the contribution of each prototypical vector in $\vect{\Omega}$ when constructing $\hat{\mathbf{z}}$. Note that the weight vector $\vect{w}$ is designed based on the similarity between the memory elements and the latent query vector $\vect{z}$ as
\begin{eqnarray}
	\label{wi}
	w_i = \frac{\exp(d(\mathbf{z},\vect{\omega}_i))}{\sum^{N_{\Omega}}_{i=1}{\exp(d(\mathbf{z},\vect{\omega}_i))}}
\end{eqnarray}
where $d(\cdot,\cdot)$ is a similarity measure, which is selected as the cosine similarity $d(\mathbf{z},\vect{\omega}_i) = \frac{\mathbf{z}^T\vect{\omega}_i}{\|\mathbf{z}\|\|\vect{\omega}_i\|}$ in this study. Hence, the memory module is designed to extract prototypical vectors in $\vect{\Omega}$ with high similarity with $\vect{z}$ to generate the representation vector $\hat{\vect{z}}$. $\hat{\vect{z}}$ will then serve as the input of the decoder to reversely map the latent domain back to the original space and generate the reconstructed signal $\hat{\vect{x}}$, i.e., $\hat{\vect{x}} =f_d ( \hat{\vect{z}},\vect{\theta}_d)$.

\subsubsection{Adversarial-training Enhanced Anomaly Detection}
To further optimize the AE, minimize potential overfitting possibility, and enlarge the gap in reconstruction error between anomalous and normal signals, we propose to leverage the GAN framework \cite{goodfellow2014generative} to incorporate adversarial training into the MemAE. The GAN consists of two modules: a generator $G$ and a discriminator $D$. $G$ is designed to generate synthetic samples that are similar to the real data, while $D$ aims to classify the data as real or artificial. The competition between $G$ and $D$ enables GAN to learn the underlying data distribution and further generate high-quality synthetic samples by optimizing the objective function:
\begin{eqnarray}
	\label{gan}
	\min_{\vect{\theta}_G}\max_{\vect{\theta}_D}\mathcal{V}(\vect{\theta}_G;\vect{\theta}_D)= \mathbb{E}_{\vect{x}\sim P_{\vect{x}}}[\log D(\vect{x};\vect{\theta}_D)] +
	\mathbb{E}_{\hat{\vect{z}}\sim P_{\vect{z}}}[\log (1- D(G(\hat{\vect{z}};\vect{\theta}_G); \vect{\theta}_D))]
\end{eqnarray}
where $\vect{\theta}_D$ and $\vect{\theta}_G$ denote the parameter sets of the discriminator and generator respectively, and $P_{\vect{x}}$ and $P_{\vect{z}}$ represent probability distributions of the real sample and latent variable spaces, respectively. The discriminator $D$ tries to maximize the objective function in Eq. (\ref{gan}) with respect to $\vect{\theta}_D$ given a fixed generator $G(\cdot; \vect{\theta}_G)$ by assigning probability of 1 to real data points $\vect{x}$ and 0 to generated samples $\hat{\vect{x}}$. The objective of generator, $G$, is to reconstruct a signal $\hat{\vect{x}}$ as similar as the original input and fool discriminator $D$ into classifying the synthetic signal as real by minimizing Eq. (\ref{gan}) with respect to $\vect{\theta}_G$ given a fixed discriminator $D(\cdot; \vect{\theta}_D)$.

In the present investigation, we adopt the MemAE as the generator of GAN, i.e., $G(\cdot)=f_d(f_e(\cdot;\vect{\theta}_e, \vect{\Omega}),\vect{\theta}_d)$, to design a Memory-Augmented Deep auto-Encoder with Generative Adversarial Network (MadeGAN) for the first-level anomaly detection. Hence, the objective function in Eq. (\ref{gan}) becomes   
\begin{eqnarray}
	\label{gan1}
	\min_{\vect{\theta}_G}\max_{\vect{\theta}_D}\mathcal{V}'(\vect{\theta}_G;\vect{\theta}_D)= \mathbb{E}_{\vect{x}\sim P_{\vect{x}}}[\log D(\vect{x};\vect{\theta}_D)+
	\log (1-D (f_d(f_e(\vect{x};\vect{\theta}_e,\vect{\Omega}),\vect{\theta}_d));\vect{\theta}_D)]
\end{eqnarray}	   
where $\vect{\theta}_G=\{\vect{\theta}_e,\vect{\Omega},\vect{\theta}_d\}$, which denotes the parameter set in MemAE. In other words, we add adversarial training into the MemAE to improve the performance in anomaly detection. Additionally, in order to guarantee the success of the training procedure, we add feature matching loss \cite{salimans2016improved} into the objective function of the generator to minimize the difference of meaningful features that are learned by hidden layers of the discriminator between the real and synthetic signals. Specifically, the feature matching loss is defined as:
\begin{eqnarray}
	\label{fm}
	\mathcal{L}_{fm} (\vect{\theta}_G) =  \mathbb{E}_{\vect{x}\sim P_{\vect{x}}}\|h_D(\vect{x}) - h_D(\hat{\vect{x}})\|
\end{eqnarray}
where $h_D(\cdot)$ denotes feature vector learned from intermediate layers of the discriminator $D$.

Moreover, in order to increase the computation efficiency, we propose to promote the sparsity of the weight vector such that the representation latent vector $\hat{\vect{z}}$ can be reconstructed with a restricted number of phenotypical patterns in the memory database. This is achieved by minimizing the L1 norm of the weight vector $\vect{w}$, i.e., 
\begin{eqnarray}
	\mathcal{L}_w(\vect{\Omega})=\|\vect{w}\|_1
\end{eqnarray}    
As such, the loss function of the MemAE is adapted by the adversarial training given a fixed discriminator $D(\cdot; \vect{\theta}_D)$ as 
\begin{eqnarray}
	\label{gen2}
	\mathcal{ L}_{edA}(\vect{\theta}_G;\vect{\theta}_D) &=& \mathbb{E}_{\vect{x}\sim P_{\vect{x}}}\|\vect{x}-\hat{\vect{x}}\|^2+\mathcal{L}_w(\vect{\Omega})+ \mathbb{E}_{\vect{x}\sim P_{\vect{x}}}\|h_D(\vect{x}) - h_D(\hat{\vect{x}})\|
\end{eqnarray}
where $\hat{\vect{x}}=f_d ( f_e(\vect{x};\vect{\theta}_e,\vect{\Omega}),\vect{\theta}_d)$. $\mathcal{ L}_{edA}(\vect{\theta}_G;\vect{\theta}_D)$ is also considered as the loss function of the generator in the proposed MadeGAN to improve the robustness and stability of network training. Thus, the overall objective of MadeGAN is given by
\begin{eqnarray}
	\min_{\vect{\theta}_G}\max_{\vect{\theta}_D}\mathcal{L}(\vect{\theta}_G,\vect{\theta}_D)=\mathcal{V}'(\vect{\theta}_G;\vect{\theta}_D)
	+ \mathcal{ L}_{edA}(\vect{\theta}_G;\vect{\theta}_D)
\end{eqnarray}

By optimizing $\mathcal{L}(\vect{\theta}_G,\vect{\theta}_D)$, the adversarial training pushes the generator (i.e., the MemAE) to improve in generating realistic reconstructions from normal inputs such that the discriminator cannot distinguish between the real and artificial signals. As such, the proposed MadeGAN is expected to enhance the anomaly detection performance of traditional AE and MemAE to better preserve the phenotypical patterns of normal ECG data, which will output low reconstruction errors for normal signals and high errors for abnormal ones.

\subsubsection{Anomaly Detection}
During the inference stage, the anomaly degree of a new query signal $\vect{x}$ is evaluated by the following anomaly score:
\begin{eqnarray}
	\label{score}
	s(\vect{x}) = \|\vect{x} - G(\vect{x})\|
\end{eqnarray}
which quantifies the discrepancy between the true signal and the signal reconstructed by patterns extracted from normal signals by $G$. The network is expected to output big anomaly scores for abnormal signals. On the other hand, a small $s(\vect{x})$ indicates that the test signal $\vect{x}$ shares similar patterns as seen in normal ECGs, which will then be classified as normal by the proposed MadeGAN.

\subsection{Transfer Learning-enhanced Second-level Classification}
The objective of second-level learning is to classify the abnormal ECGs identified from first-level MadeGAN into different types of arrhythmias for accurate disease detection, as shown in Fig. \ref{Fig:2levelAD}(b). Note that the performance of deep learning-based classification depends, to a great extent, on the quality and quantity of training data. However, due to the expensive data collection and labeling process, data associated with abnormal heart conditions is significantly less than that from the healthy condition. Moreover, the occurrence rates of different diseases are highly diverse, depending on the underlying characteristics of the patient's health condition, which leads to imbalanced data sizes for different disease types. Traditional predictive modeling based on such small and imbalanced data  tends to generate biased estimates, leading to unsatisfactory detection performance. As such, innovative design of network architecture and novel training technique are urgently needed to study ECG signals for accurate disease identification. 

Here, we propose to take advantage of the large volume of normal ECG cycles to improve the second-level classification through transfer learning \cite{pan2009survey}. Note that even though the exact distribution and morphological shapes between normal and abnormal ECGs are different, they share common patterns on the macroscale. For example, each ECG consists of a P-wave, QRS-complex, and T-wave. Hence, such macroscale pattern information in normal ECGs can be transferred to the second-level learning to improve the classification performance.

\begin{figure}
	\begin{center}
		\includegraphics[height=3.5in]{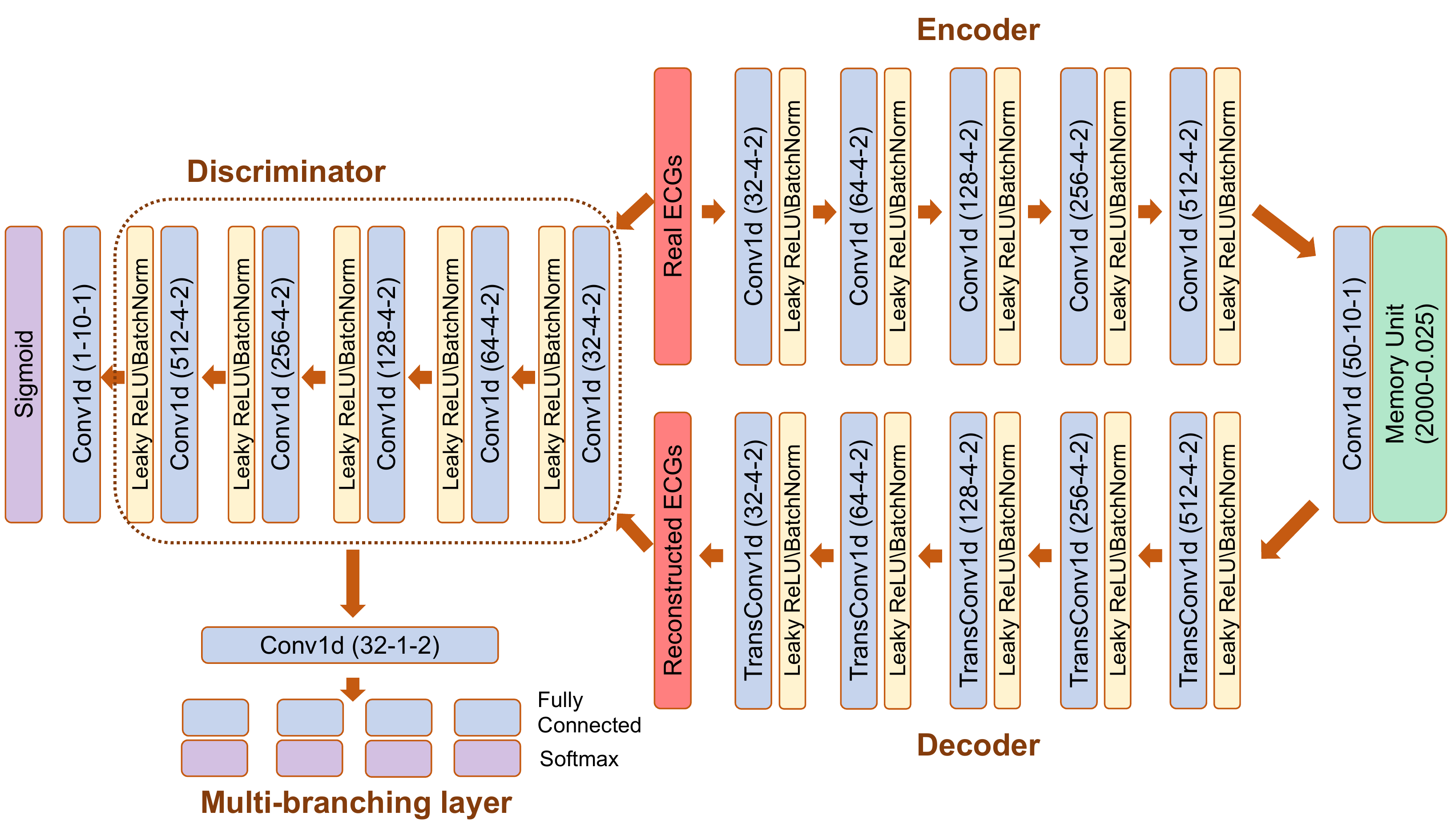}
		\caption{The architecture details of our proposed hierarchical deep learning framework.}
		\label{Fig:model_detail}
	\end{center}      	
\end{figure}

Transfer learning generally involves two different domains: source domain $\mathcal{X}_S$, and target domain $\mathcal{X}_T$, and the corresponding two different learning tasks $\mathcal{T}_S$ and $\mathcal{T}_T$, respectively. The objective of transfer learning is to leverage the knowledge learned from $\mathcal{T}_S$ from the source domain $\mathcal{X}_S$ to improve the learning task $\mathcal{T}_T$ in the target domain $\mathcal{X}_T$. Specifically, in the present investigation, the source domain $\mathcal{X}_S$ consists of the  normal ECG data and the associated task $\mathcal{T}_S$ is anomaly detection. Our target task $\mathcal{T}_T$ for second-level learning is multi-class classification to identify different types of arrhythmias from the data in the target domain $\mathcal{X}_T$ (i.e., abnormal ECGs). In order to achieve the knowledge transfer, we adopt the discriminator ${D}$ in the MadeGAN with learned network parameters $\vect{\hat{\theta}}_D$ from the first-level training to improve the learning of the multi-class classification $\mathcal{T}_T$ in the second level. Note that we remove the last 1D CNN and fully connected layers from discriminator $D$ (see more detail in Fig.\ref{Fig:model_detail}.)  and use the remaining network structure ${\bar{D}(\cdot;\vect{\hat{\theta}}_D)}$ as the feature extractor to extract critical patterns from the abnormal ECG data, which will then be fed into a shallow classifier (i.e., 1D CNN) for different arrhythmia identification. Note that ${\bar{D}(\cdot;\hat{\theta}_D)}$ is frozen during the whole second-level training process (i.e, $\vect{\hat{\theta}}_D$ is kept unchanged).

In addition, because the common problem of imbalanced data may still exist in different types of abnormal signals, we add the multi-branching (MB) architecture \cite{wang2021multi} as shown in Fig. \ref{Fig:2levelAD}(c) to the 1-D CNN to address the imbalanced data issue and further improve the classification performance. We denote the  the abnormal dataset as $\mathcal{D}_A=\{\vect{X}_A,\vect{y}_A\}$, where $\vect{X}_A$ is the set of abnormal ECGs and $\vect{y}_A$ denotes the corresponding label set. Assume there are $M$ different types of arrhythmia or $\mathcal{D}_A$ consists of $M$ different sub-groups, i.e., $\mathcal{D}_A=\{\mathcal{D}_1,\dots,\mathcal{D}_M\}$, and assume class $l$ contains the smallest data samples, i.e., $|\mathcal{D}_l|\leq |\mathcal{D}_m|$ for $m\neq l$ and $m\in\{1,\dots,M\}$. We generate $N_b$ balanced datasets, $\mathcal{D}_{Ai}$'s ($i\in\{1,\dots,N_b\}$), from $\mathcal{D}_A$ by under-sampling the class with larger sample size, i.e., $\mathcal{D}_{Ai}=\{\mathcal{D}_{1i},\dots,\mathcal{D}_{(l-1)i},\mathcal{D}_{l},\mathcal{D}_{(l+1)i},\dots,\mathcal{D}_{Mi}\}$, where $\mathcal{D}_{mi}$ is a subset by under-sampling the data $\mathcal{D}_{m}$ from type $m$ arrhythmia ($m\in\{1,\dots,l-1,l+1,\dots,M\}$) to the same size with $\mathcal{D}_l$. The corresponding $N_b$ branching outputs for the $N_b$ balanced datasets are further created and attached to the 1D CNN for robust and accurate disease identification. 

The shallow 1D CNN will be trained by all the $N_b$ datasets, and each branching output will be trained by the corresponding balanced sub-dataset. This MB architecture and training procedure will enable the network to learn more effectively from the imbalanced dataset for reliable classification \cite{wang2021multi}. The objective of second-level learning is to search a set of optimal parameters such that the network outputs a high probability for the true label and low probabilities for others. As such, we select the multi-class cross-entropy as the loss function of second-level network:
\begin{eqnarray}
	\mathcal{L}(\vect{\theta}_{2nd};\vect{X}_A)=-\sum_{p=1}^{N_A}\sum_{i=1}^{N_b}\mathcal{I}(p\in\mathcal{D}_{Ai}) \Big(\sum_{m=1}^{M}y_p^{(m)}\log(\hat{P}_i^{(m)}(\vect{\theta}_{2nd};\bar{D}(\vect{x}_A^p;\vect{\hat{\theta}}_D)))\Big)
	\label{Eq: obj}
\end{eqnarray}
where $\vect{\theta}_{2nd}$ denotes the parameter set of the second-level network, $N_A$ is the sample size of the abnormal dataset, $\mathcal{I}(\cdot)$ is an indicator function, and $\hat{P}_i^{(m)}(\vect{\theta}_{2nd};\bar{D}(\vect{x}_A^p;\vect{\hat{\theta}}_D))$ is the predicted probability of type $m$  arrhythmia from branching output $i$ given the features $\bar{D}(\vect{x}_A^p;\vect{\hat{\theta}}_D))$ extracted by the discriminator from the input $\vect{x}_A^p$. Please see more details about the MB network in our prior work \cite{wang2021multi}. The final predicted probability of class $m$ for input $\vect{x}_A^p$ is given by
\begin{eqnarray}
	\hat{P}^{(m)} =\frac{1}{N_b}\sum_{i=1}^{N_b} \hat{P}_i^{(m)}(\vect{\theta}_{2nd};\bar{D}(\vect{x}_A^p;\vect{\hat{\theta}}_D))
\end{eqnarray}

Integrating the MB structure and the discriminator-based transfer learning with the 1D CNN enables our second-level model (MB-Conv1d-Discrim) to overcome the data-lacking and imbalanced data issues, and further improve the classification performance for accurate disease identification.

\section{Experimental Design and Results}
\label{sec4} 
\subsection{Data Description and Experimental Design}  
The dataset used in this study is obtained from MIT-BIH database \cite{moody2001impact}. This dataset contains 48 half-hour  ECG recordings, which is obtained from 47 subjects from the Beth Israel
Hospital Arrhythmia Laboratory between 1975 and 1979. Each recording is digitized at 360 samples per second with 11-bit resolution. Two or more independent cardiologists annotate the beats in each recording. We remove four recordings (102, 104, 107, 217) because of their poor signal quality. Each heartbeat is segmented through the following procedure: (1) The Pan-Tompkins QRS detection algorithm \cite{pan1985real} is first employed to detect all R-peak locations in original ECG recordings; (2) 320 samples are then selected for each heartbeat with 140 samples before the corresponding R-peak and 180 samples after it. This procedure guarantees that segmented signals have the same dimension and approximately cover one heart cycle. In total, our dataset contains 97,553 annotated beats including: 86,717 normal signals (N), 7,008 premature ventricular contractions (V), 3,026 supraventricular premature beats (S), and 802 ventricular fusion beats (F). Additionally, we employ the high-pass finite impulse response (FIR) filter to reduce multiple types of noise commonly seen in ECG signals such as power-line interference and baseline wandering \cite{an2020comparison}.

The performance of our first-level MadeGAN in anomaly detection will be compared with a pure AutoEncoder (AE), AE with memory module (MemAE), and BeatGAN \cite{zhou2019beatgan}. The second-level model (i.e., MB-Conv1d-Discrim) will be benchmarked with a pure 1D CNN (Conv1d), 1D CNN with the generator feature extractor (Conv1d-Gene), 1D CNN with the encoder feature extractor (Conv1d-Encoder), 1D CNN with the discriminator feature extractor (Conv1d-Discrim), 1D CNN with an MB layer (MB-Conv1d), Conv1d-Gene with an MB layer (MB-Conv1d-Gene),  and Conv1d-Encoder with an MB layer (MB-Conv1d-Encoder). The performance of both the first- and second-level learning will be evaluated according to:  two overall performance metrics, i.e., Area Under Receiver-Operating-Characteristic Curve (AUROC), and Area Under Precision-Recall Curve (AUPRC),  and four point performance metrics, i.e., $Recall$, $Precision$, $f$-score, and $Accuracy$.  The ROC characterizes the relationship between the False Positive Rate ($FPR$) and True Positive Rate ($TPR$), and AUROC is a score quantifying the general prediction performance across all thresholds. Similarly, the AUPRC evaluates the overall relationship between $Recall$ and $Precision$. 

To evaluate the performance metrics for the first-level anomaly detection, we calculate the anomaly score, $s_i$, for each sample in the test set, generating the set $S_A=\{s_1,s_2,\dots,s_{N_T}\}$, where $N_T$ denotes the number of samples in the test set. Then, we apply feature scaling to convert the anomaly scores into the probabilistic range of [0,1]:
\begin{eqnarray}
	\label{scale}
	s_i^{'} = \frac{s_i - \min(S_A)}{\max(S_A) - \min(S_A)} 
\end{eqnarray}
As such, different thresholds on the normalized score $s_i^{'}$ will be applied to calculate $FPR$, $TPR$, $Recall$, and $Precision$, and further compute the AUROC and AUPRC scores for the first-level anomaly detection.

\begin{figure}
	\begin{center}
		\includegraphics[width=5in]{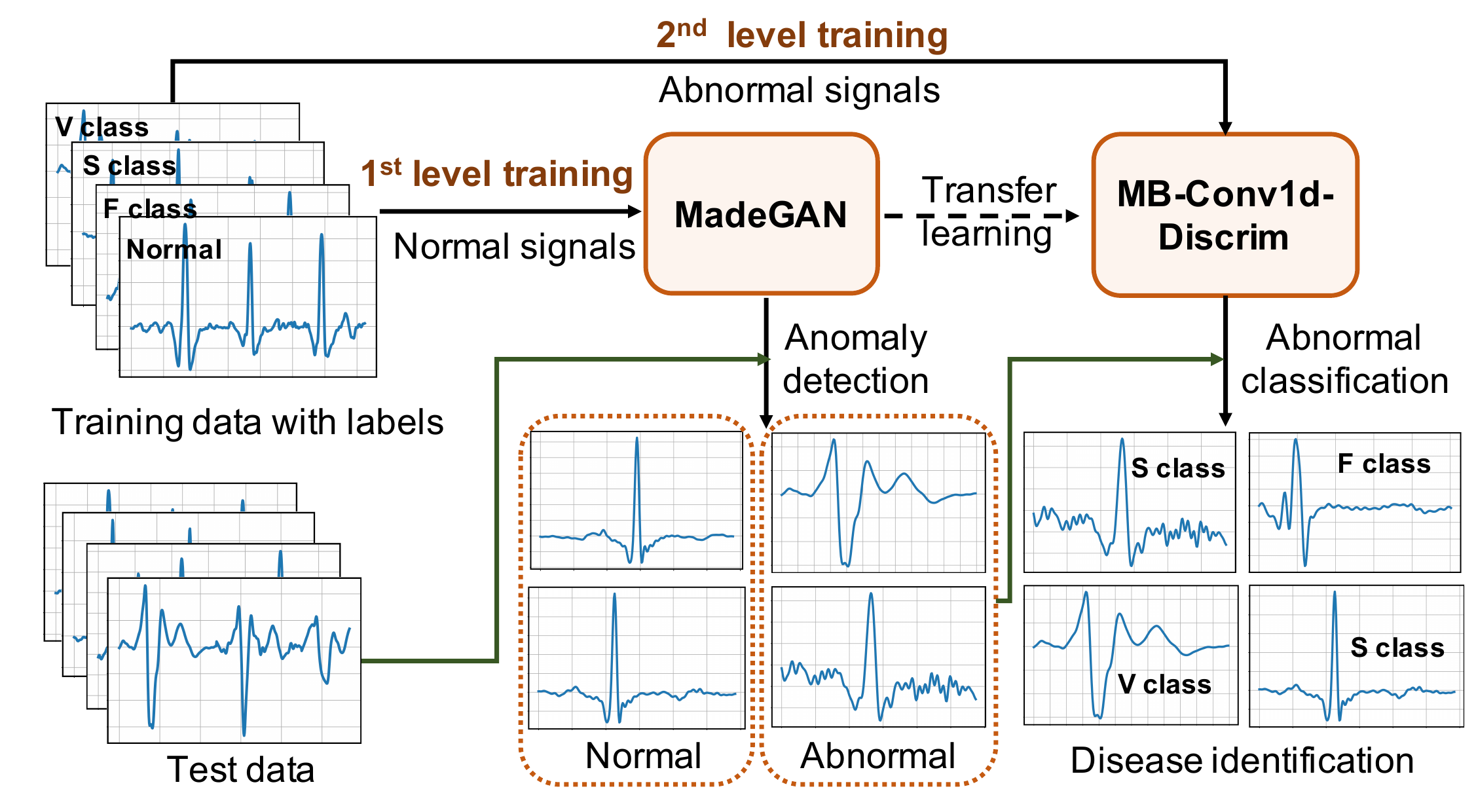}
		\caption{Training and testing procedures for the proposed hierarchical deep learning framework.}
		\label{Fig:train_test}
	\end{center}      	
\end{figure}

\subsection{Training and Testing Procedures}
Fig. \ref{Fig:model_detail} shows the detailed architecture of the proposed two-level framework. Our first-level MadeGAN contains four core modules: encoder, decoder, memory module, and discriminator. We use five 1D convolutional (Conv1d) layers in the encoder. Note that the notation of $(n_{filter}, n_{size}, n_{stride})$ represents that there are $n_{filter}$ filters with a filter size of $n_{size}$ and the stride of $n_{stride}$. For example, (32-4-2) means that this Conv1d layer is constructed by 32 filters with the filter size of 4 and stride of 2. The latent space dimension of the encoder is 50, i.e., $d_{\vect{z}}=50$. Additionally, we use batch-normalization and leaky ReLU activation to each Conv1d layer to make the model training more stable. The decoder is a mirrored version of the encoder with five transposed Conv1d layers. The memory model contains 2000 prototypical vectors of normal ECGs, i.e., $N_{\Omega}=2000$. The discriminator has the same core architecture as the encoder with an additional Conv1d and fully connected layers for classification. Our second-level model consists of a Conv1d layer, an MB layer with 4 branching outputs, and a feature extractor. Note that the feature extractor is composed of the discriminator without the last Conv1d and fully connected layers (i.e., the part of discriminator inside the dashed box in Fig. \ref{Fig:model_detail}). Specifically, the abnormal signal will first go through Conv1d (32-4-2), Conv1d (64-4-2), Conv1d (128-4-2), Conv1d (256-4-2), and Conv1d (512-4-2) layers in the discriminator including all Leaky Relu and Batch normalized layers. Then, the output from Conv1d (512-4-2) layer will go to the shallow 1D CNN layer (i.e., Conv1d (32-1-2)) and an MB layer to make the final second-level prediction. Adam optimizer with an initial rate of $lr = 0.0002$ and momentums of $\beta_{1} = 0.5$, $\beta_{2}=0.999$ is used for network training. The network hyper-parameters, i.e., $\{n_{filter}, n_{size}, n_{stride}, d_z, N_{\Omega}, lr, \beta_{1}, \beta_{2}\}$ are selected by empirical fine-tuning.

Fig. \ref{Fig:train_test} shows the training and testing procedures for the proposed model. Specifically, during the first-level MadeGAN training, we use 90\% normal signals as the training set and the rest of 10\% normal signals as the test set. In order to evaluate the anomaly detection performance of MadeGAN in the testing phase, all abnormal ECGs are combined with the rest 10\% normal ones to form the test set. During the second-level training, we use 90\% abnormal ECGs as the training set, which contains 3 different types of arrhythmias (i.e., the V, S, and F samples). The rest of 10\% abnormal signals is used as the test set to evaluate the performance of the second-level classification. Note that during the second-level training, the parameter $\vect{\hat{\theta}}_D$ of discriminator ${D}$ adopted from MadeGAN is held frozen to achieve the knowledge transfer from the first-level learning.

\subsection{Experimental Results}
\subsubsection{First-level Anomaly Detection}

Table \ref{MadeGAN_score} shows the performance comparison between MadeGAN and other benchmarks in the first-level anomaly detection. Note that MadeGAN achieves the best performance with AUROC and AUPRC scores of 0.954 and 0.936. Specifically, MadeGAN improves on AUROC by 5.3\% compared with the pure AE (0.906),  by 2.9\% compared with MemAE (0.927), and by 1.4\% compared with BeatGAN (0.941). Furthermore, the improvement is more significant in terms of AUPRC: with the improvement of 5.6\%, 3.8\%, and 1.5\% compared with the pure AE (0.886), MemAE (0.902), and BeatGAN (0.922), respectively.

\begin{table}[h!]
	\caption{The comparison of AUROC and AUPRC scores of our MadeGAN and other models in the first-level anomaly detection.}
	\centering
	{\scriptsize
		{
			\setlength{\tabcolsep}{2mm}{
				\begin{tabular}{c|cccccc}
					\toprule
					& \makecell[c]{AUROC}
					&\makecell[c]{AUPRC}\\
					\hline \\[-0.95em]         			
					\makecell[c]{MadeGAN} &0.954&0.936
					\\
					\makecell[c]{BeatGAN}& 0.941&0.922
					\\
					\makecell[c]{MemAE}& 0.927&0.902
					\\
					\makecell[c]{AE}& 0.906&0.886
					\\
					\bottomrule
				\end{tabular}
	}}}
	\label{MadeGAN_score}
\end{table}      

\begin{figure}
	\begin{center}
		\includegraphics[width=6in]{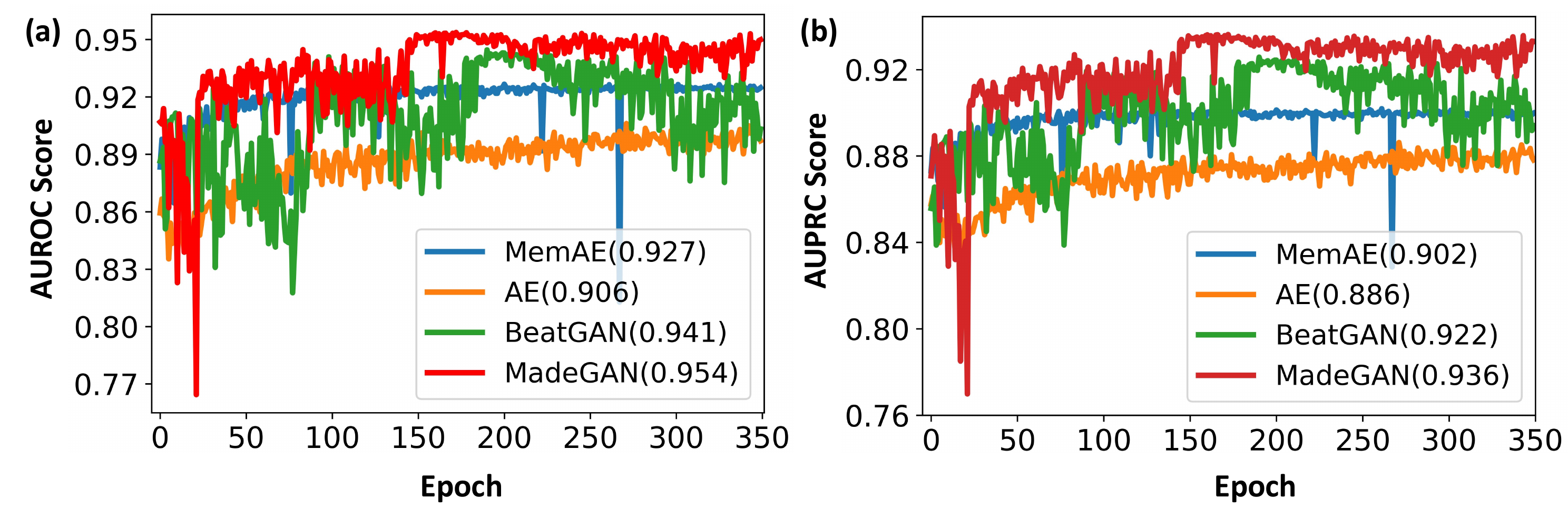}
		\caption{The variations of (a) AUROC score and (b) AUPRC score over epochs in test set for different methods during the first-level anomaly detection.}
		\label{Fig:MadeGAN_score}
	\end{center}      	
\end{figure}

\begin{figure}
	\begin{center}
		\includegraphics[width=5.5in]{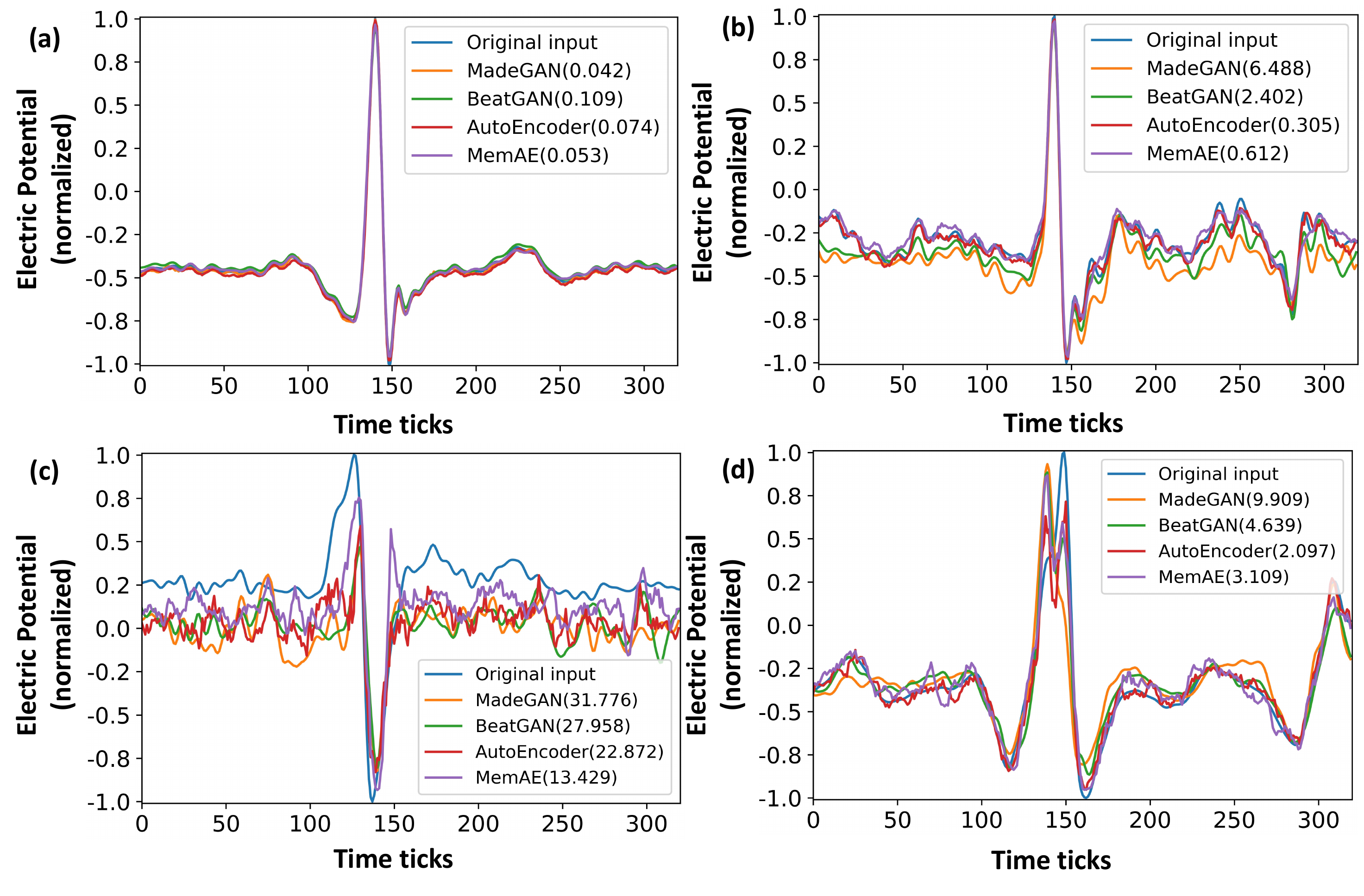}
		\caption{Reconstructed ECG signals by the proposed MadeGAN and other benchmarks for (a) N class (b) S class (c) V class (d) F class.}
		\label{Fig:recon_signal}
	\end{center}      	
\end{figure}

Fig. \ref{Fig:MadeGAN_score} (a) and (b) display the AUROC and AUPRC over 350 epochs in the test dataset for different methods. Note that the pure AE and MemAE converge in the first 100 epochs but achieve relatively low performance scores, i.e., AUROC of 0.906 and 0.927, and AUPRC of 0.886 and 0.902, respectively. On the other hand, the performance scores of MadeGAN and BeatGAN improve over epochs and achieve better AUROC of 0.954 and 0.941, and AUPRC of 0.936 and 0.922, respectively. This is due to the fact that both MadeGAN and BeatGAN incorporate the adversarial training into the reconstruction, which alleviates the potential overfitting problem in AE training. In addition, the memory module enables MadeGAN to effectively account for the possible variations in normal signals and reduce the possibility of reconstructing abnormal ECGs with small errors. This further enlarges the gap in reconstruction errors between normal and abnormal signals, which makes MadeGAN achieve better performance compared with BeatGAN. Table \ref{point_metrics} further shows the performance comparison in terms of point metrics. Note that our MadeGAN yields the best $Recall$, $Precision$, $f$-score, and $Accuracy$ of 0.954, 0.856, 0.902, and 0.885 respectively among the four methods.

\begin{table}[h!]
	\caption{The comparison of point metrics of MadeGAN and other methods in the first-level anomaly detection.}
	\centering
	{\scriptsize
		{
			\setlength{\tabcolsep}{2mm}{
				\begin{tabular}{c|cccccc}
					\toprule
					& \makecell[c]{Recall}
					&\makecell[c]{Precision}
					&\makecell[c]{$f$-score}
					&\makecell[c]{Accuracy}\\
					\hline \\[-0.95em]         			
					\makecell[c]{MadeGAN} &0.954&0.856&0.902&0.885
					\\
					\makecell[c]{BeatGAN}& 0.944&0.826&0.881&0.884
					\\
					\makecell[c]{AE}& 0.894&0.760&0.821&0.784
					\\
					\makecell[c]{MemAE}&0.886&0.782&0.831&0.800
					\\
					\bottomrule
				\end{tabular}
	}}}
	\label{point_metrics}
\end{table}

Fig. \ref{Fig:recon_signal} displays the examples of reconstructed ECG signals generated by our first-level MadeGAN and other benchmarks for normal ECGs and different types of abnormal ECGs. Here, we adopt the reconstruction error (i.e. $\|\hat{\vect{x}}-\vect{x}\|^2$) to evaluate the performance for different approaches. Because all models are trained purely by normal ECGs, we expect that a small reconstruction error will be produced from the normal signal and a large error will be generated from the abnormal signal. According to Fig.\ref{Fig:recon_signal} (a), although the reconstructed signals from all the methods match with the input normal signal closely, our MadeGAN generates the smallest reconstruction error (0.042) compared to other approaches (0.109 for BeatGAN, 0.074 for AE, and 0.053 for MemAE). On the other hand, according to Fig. \ref{Fig:recon_signal} (b)-(d), MadeGAN is able to generate the most dissimilar reconstructions and produces the largest reconstruction errors for abnormal signals. For example, the reconstruction error generated by MadeGAN  for the S class is 6.488, which is significantly bigger than 2.402 by BeatGAN, 0.305 by AE, and 0.612 by MemAE. 


\subsubsection{Second-level Arrhythmia Identification}


\begin{table}[h!]
	\caption{The comparison of AUROC and AUPRC between MB-Conv1D-Discrim and other approaches in the second-level classification}
	\centering
	{\scriptsize
		{
			\setlength{\tabcolsep}{1mm}{
				\begin{tabular}{c|cccccc}
					\toprule
					\makecell[c]{}&\multicolumn{2}{c}{S Class}&\multicolumn{2}{c}{V Class}&\multicolumn{2}{c}{F Class}\\
					\hline \\[-0.95em]         			
					\makecell[c]{} &AUROC&AUPRC&AUROC&AUPRC&AUROC&AUPRC\\
					\hline \\[-0.95em] 
					\makecell[c]{MB-Conv1D-Discrim}& 0.988&0.989&0.988&0.987&0.997&0.988\\
					\makecell[c]{MB-Conv1D-Encoder}& 0.963&0.955&0.960&0.938&0.952&0.899\\
					\makecell[c]{MB-Conv1D-Gene}& 0.983&0.981&0.985&0.938&0.952&0.953\\
					\makecell[c]{MB-Conv1D}& 0.987&0.987&0.984&0.975&0.981&0.939\\
					\makecell[c]{Conv1D-Discrim}& 0.985&0.981&0.984&0.987&0.963&0.911\\
					\makecell[c]{Conv1D-Encoder}& 0.974&0.956&0.979&0.986&0.952&0.719\\
					\makecell[c]{Conv1D-Gene}& 0.979&0.961&0.980&0.981&0.959&0.828\\
					\makecell[c]{Conv1D}& 0.976&0.953&0.974&0.983&0.924&0.534\\
					
					\bottomrule
				\end{tabular}
	}}}
	\label{MadeGAN_AUROC_AUPRC}
\end{table}

{Table. \ref{MadeGAN_AUROC_AUPRC} shows the AUROC and AUPRC of our MB-Conv1d-Discrim and other MB-based and Conv1D-based benchmarks in the second-level arrhythmia classification.  Note that a good model will have a large AUROC and AUPRC. According to Table. \ref{MadeGAN_AUROC_AUPRC}, our MB-Conv1d-Discrim has the best AUROC and AUPRC among other  methods. Specifically, compared with MB-Conv1d-Gene and MB-Conv1d-Encoder, our MB-Conv1d-Discrim increases AUROC from 0.983 and 0.963 to 0.988, and increases AUPRC from 0.981 and 0.955 to 0.989 for the S class. Similarly, for the V class, MB-Conv1d-Discrim improves AUROC from 0.985 and 0.960 to 0.988, and increases AUPRC from 0.981 and 0.938 to 0.987 compared with MB-Conv1d-Gene and MB-Conv1d-Encoder. The improvement is more significant in the F class, which has the smallest sample size among the three classes. Specifically, MB-Conv1d-Discrim achieves 1.5\% and 3.8\% performance improvement on AUROC, and 3.8\% and 10.0\% improvement on AUPRC compared with MB-Conv1d-Gene and MB-Conv1d-Encoder for the F class. 
	
	Similarly, MB-Conv1d-Discrim increases AUROC from 0.979 and 0.974 to 0.988, and improves AUPRC from 0.961 and 0.956 to 0.989 compared with Conv1d-Gene and Conv1d-Encoder for the S class. For the V class, MB-Conv1d-Discrim enhances AUROC from 0.980 and 0.979 to 0.984, and increases AUPRC from 0.981 and 0.986 to 0.987 compared with Conv1d-Gene and Conv1d-Encoder. For the F class, MB-Conv1d-Discrim achieves 3.9\% (from 0.959 to 0.997) and 4.7\% (from 0.952 to 0.997) performance improvement on AUROC compared with Conv1d-Gene and Conv1d-Encoder, and achieves 19.3\% (from 0.828 to 0.988) and 37.4\% (from 0.719 to 0.988) improvement on AUPRC. More importantly, our MB-Conv1D-Discrim achieves a significant improvement of 7.3\% on AUROC (from 0.924 to 0.997) and 85\% on AUPRC (from 0.534 to 0.988) for F class compared with the pure Conv1D without any transfer learning or MB layer. In order to further show that our proposed MB-Conv1d-Discrim dominates other method, we plot the Detection Error Tradeoff Curves (DETC) with logarithmic x- and y-axis in Appendix. Please see the details in Appendix.

	According to Table \ref{MadeGAN_AUROC_AUPRC}, the discriminator-based feature extractor outperforms the generator-based and encoder-based feature extractors. This is due to the fact that the discriminator is trained to classify the signal as fake or real. In other words, classification is the learning task of the discriminator in the first-level training, which is similar to the second-level task (i.e., multi-class classification). However, the learning tasks for the generator or the encoder are signal reconstruction and latent-space generation respectively, which are different from the second-level learning task. As such, the discriminator-based feature extractor yields the best performance to improve the second-level classification.


	\begin{table}[h!]
		\caption{Confusion matrix on the test dataset from the second-level MB-Conv1d-Discrim method.}
		{\renewcommand{\arraystretch}{2}
			\setlength{\tabcolsep}{10mm}{
				\begin{tabular}{l|l|c|c|c}
					\multicolumn{2}{c}{}&\multicolumn{2}{c}{Predicted labels}&\\
					\cline{3-5}
					\multicolumn{2}{c|}{}&S&V&\multicolumn{1}{c|}{F}\\
					\cline{2-5}
					\multirow{2}{*}{\makecell[c]{True\\labels}}& S & $299$ & $4$ &\multicolumn{1}{c|}{0}\\
					\cline{2-5}
					& V & $4$ & $687$ &\multicolumn{1}{c|}{10}\\
					\cline{2-5}
					& F & $0$ & $1$ &\multicolumn{1}{c|}{80}\\
					\cline{2-5}
				\end{tabular}
		}}
		\label{confusion_matrix}
	\end{table}  
	
	It is also worth noting that MB-based methods outperform non-MB-based methods. This improvement is more significant for the class with a small sample size (i.e., F class). Specifically, as shown in Table. \ref{MadeGAN_AUROC_AUPRC}, MB-Conv1d increases AUROC and AUPRC from 0.924 to 0.981 and from 0.534 to 0.939 respectively, compared with the pure Conv1d model. MB-Conv1d-Discrim increases AUROC and AUPRC from 0.963 to 0.997 and from 0.911 to 0.988 respectively, compared with the Conv1d-Discrim. This is due to the fact that the MB architecture guarantees each branching output is trained by a balanced sub-dataset to effectively relieve the imbalanced data issue. As such, the integration of transfer learning and MB architecture has significantly improved classification performance in the second-level learning.}

Table \ref{confusion_matrix} further shows the confusion matrix on the test dataset provided by the proposed second-level model (i.e., MB-Conv1d-Discrim). Note that even for the F class, our model provides pretty good classification performance. Specifically, 80 F samples are correctly predicted out of the 81 total F samples. Table \ref{point_metrics2} shows the performance comparison of point metrics between MB-Conv1d-Discrim and other benchmarks. Note that MB-Conv1d-Discrim yields the best $Recall$, $Precision$, $f$-score, and $Accuracy$ of 0.976, 0.976, 0.975, and 0.976,  respectively. Note that MB-based methods also outperform non-MB methods in terms of the point metrics. For example, MB-Conv1d-Discrim, MB-Conv1d-Encoder, MB-Conv1d-Generator, and MB-Conv1d achieve 2.5\%, 7.3\%, 4.7\%, and 22.3\% improvement on the $f$-score compared with their non-MB counterparts. It is also worth noting that MB-Conv1d-Discrim achieves a more significant improvement on $Recall$, $Precision$, and $f$-score than $Accuracy$ compared to the pure Conv1d without any transfer learning or MB layer. Specifically, MB-Conv1d-Discrim achieves 29.8\% performance improvement on $Recall$, 21.8\% on $Precision$, and 20.5\% on $f$-score. This is due to the fact that $Recall$, $Precision$, and $f$-score are more robust evaluation metrics when there exists an imbalanced issue, which can be effectively addressed by the MB architecture and transfer learning.


\begin{table}[h!]
	\caption{The comparison of point metrics between MB-Conv1d-Discrim and other approaches in the second-level classification}
	\centering
	{\scriptsize
		{
			\setlength{\tabcolsep}{2mm}{
				\begin{tabular}{c|cccccc}
					\toprule
					& \makecell[c]{Recall}
					&\makecell[c]{Precision}
					&\makecell[c]{$f$-score}
					&\makecell[c]{Accuracy}\\
					\hline \\[-0.95em]         			
					\makecell[c]{MB-Conv1d-Discrim} &0.976&0.976&0.975&0.976
					\\
					\makecell[c]{MB-Conv1d-Encoder}& 0.881&0.881&0.879&0.891
					\\
					\makecell[c]{MB-Conv1d-Generator}& 0.913&0.933&0.922&0.928
					\\
					\makecell[c]{MB-Conv1d}& 0.931&0.936&0.934&0.939
					\\
					\makecell[c]{Conv1d-Discrim} &0.955&0.947&0.951&0.973
					\\
					\makecell[c]{Conv1d-Encoder}& 0.786&0.869&0.819&0.917
					\\
					\makecell[c]{Conv1d-Generator}& 0.888&0.877&0.881&0.934
					\\
					\makecell[c]{Conv1d}& 0.752&0.801&0.770&0.896
					\\
					\bottomrule
				\end{tabular}
	}}}
	\label{point_metrics2}
\end{table}  

\subsubsection{Comparison Study with Existing Literature}
In order to benchmark with existing literature, we use the 5-fold cross validation to further evaluate the proposed first- and second-level models. The 5-fold cross validation has been proved to be able to generate a less biased estimate of model performance and is widely utilized to evaluate machine learning models. Table \ref{literature} summarizes the comparison of AUROC and AUPRC between the first-level MadeGAN and existing methods that used the MIT-BIH dataset for ECG anomaly detection. Our MadeGAN yields the average of the 5 folds $\pm$ 1 standard deviation (std) for AUROC and AUPRC of 0.950$\pm$0.002 and 0.922$\pm$0.003, respectively, which outperforms existing methods \cite{zhou2019beatgan,shin2020decision}. 

\begin{table}[h!]
	\caption{The comparison of AUROC and AUPRC between MadeGAN and existing anomaly detection models.}
	\centering
	{\scriptsize
		{
			\begin{tabular}{c|ccc}
				\toprule
				\makecell[c]{Authors}&\makecell[c]{Methods}&\makecell[c]{AUROC}&\makecell[c]{AUPRC}\\
				\hline \\[-0.95em]         			
				\makecell[c]{Wang et al.} &MadeGAN&0.950&0.922\\
				\makecell[c]{Zhou et al.,\cite{zhou2019beatgan}, 2019}& BeatGAN&0.945&0.911\\
				\makecell[c]{Shin et al.\cite{shin2020decision}, 2019} &AnoGAN&0.948&-\\
				\bottomrule
			\end{tabular}
	}}
	\label{literature}
\end{table}

\begin{table}[h!]
	\caption{The comparison of performance scores between our MB-Conv1d-Discrim and existing classification models.}
	\centering
	{\scriptsize
		{
			\resizebox{.5\textwidth}{!}{%
				\begin{tabular}{c|ccccccc}
					\toprule
					\makecell[c]{Authors}&\makecell[c]{AUROC}&\makecell[c]{AUPRC}&\makecell[c]{Recall}&\makecell[c]{Precision}&\makecell[c]{$f-$score}&\makecell[c]{Accuracy}\\
					\hline \\[-0.95em]         			
					\makecell[c]{Wang et al.} &0.989&0.984&\makecell[c]{0.964}&\makecell[c]{0.967}&\makecell[c]{0.965}&\makecell[c]{0.967}\\
					\makecell[c]{Li et al.,\cite{li2016ecg}}&-&-&\makecell[c]{0.647}&\makecell[c]{0.475}&\makecell[c]{-}&\makecell[c]{0.946}\\
					\makecell[c]{Acharya et al.\cite{acharya2017deep}}&-&-&\makecell[c]{0.960}&\makecell[c]{0.915}&\makecell[c]{-}&\makecell[c]{0.940}\\
					\makecell[c]{Mousavi et al.\cite{mousavi2021ecg}}&-&-&\makecell[c]{0.831}&\makecell[c]{0.923}&\makecell[c]{-}&\makecell[c]{0.987}\\
					\makecell[c]{Kachuee et al.\cite{kachuee2018ecg}}&-&-&\makecell[c]{-}&\makecell[c]{-}&\makecell[c]{-}&\makecell[c]{0.934}\\	
					\bottomrule
			\end{tabular}}%
			
	}}
	\label{literature2}
\end{table}

Table \ref{literature2} summarizes the comparison of the performance scores between our second-level model (i.e., MB-Conv1d-Discrim) and existing approaches. Our method yields the average ($\pm$1 std) of the 5 folds for AUROC of 0.989($\pm$0.001), AUPRC of 0.984($\pm$0.002), $Recall$ of 0.964($\pm$0.006), $Precision$ of 0.967($\pm$0.005), and $f$-score of 0.965($\pm$0.005), which significantly outperforms existing methods. Note that the $Accuracy$ of our model (i.e., 0.967 $\pm$ 0.005) is slightly smaller than the accuracy (i.e., 0.987) in \cite{mousavi2021ecg}. However, the $Accuracy$ alone may not serve as a reliable performance measure when evaluating machine learning models with imbalanced data issue \cite{flach2015precision}. For example, if we have a dataset with 10,000 negative samples and 10 positive samples, we can obtain an $Accuracy$ as high as 99.9\% if we train a model to predict all samples as negative. However, this model is meaningless because we are more interested in the ability of the model to predict positive samples, which are commonly evaluated by $Recall$ and $Precision $. Note that our model has significantly better $Recall$ and $Precision$ (i.e., 0.964 and 0.967) compared with  \cite{mousavi2021ecg} (i.e., 0.831 and 0.923).

\subsection{Discussion on the Limitation}

One potential limitation of this work is that we only focus on single-lead ECG analysis. Although single-lead ECG analysis has great application potential for wearable everyday health monitoring, cardiologists usually make the medical diagnosis based on multi-lead ECG signals (e.g., 12-lead ECGs) from patients with severe heart disease. It is worth noting that 12-lead ECGs are multi-channel time-series signals, which can be considered as 3D tensors characterized by $(n_b, n_c, n_t)$, where $n_b$ is the batch size, $n_c$  is the number of channels ($n_t=12$ for 12-lead ECGs), and $n_t$ is the number of time ticks of one heartbeat. By re-designing the input layer to incorporate the channel information, the 3D tensor data can be fed to and train the network for anomaly detection and arrhythmia identification. However, the predictive performance depends to a great extent on the network structure details and hyperparameter selection. One of our future research directions will focus on adapting the proposed hierarchical deep learning framework and designing effective network structures to capture both the temporal dynamics within each channel and the correlations across different channels to analyze multi-channel ECG signals for heart disease detection.

Another potential limitation is that the finite impulse response (FIR) filter for ECG denoising may not be effective to remove motion artifacts \cite{satija2018review,satija2017automated,li2014machine}. In order to further increase the applicability of the proposed method in real clinical diagnosis, more advanced signal quality assessment methods are needed in preprocessing step \cite{satija2018review,satija2017automated,li2014machine}, which will be investigated and combined with our model for ECG signal analysis in our future work.

\section{Conclusions}
\label{sec5} 
In this paper, we develop a two-level hierarchical deep learning framework with Generative Adversarial Network to investigate ECG data for robust and reliable identification of heart diseases. In the first-level learning, we propose a Memory-Augmented Deep auto-Encoder with Generative Adversarial Network (MadeGAN) to achieve anomaly detection (i.e., binary classification of normal and abnormal signals). In the second-level learning, we employ transfer learning by adopting the discriminator with learned network parameters from the first-level training to handle the data-lacking problem and achieve the multi-class classification among different types of arrhythmias. Additionally, the multi-branching technique is used in the second-level model to handle the imbalanced data issue and enhance the classification performance. Experimental results show that our framework effectively captures the disease-altered feature patterns from ECG signals, yielding better performance in predicting heart disease with higher performance scores compared with existing methods. Moreover, this hierarchical deep learning framework can be broadly implemented to study other waveform data such as electroencephalography (EEG) and photoplethysmography (PPG) for smart anomaly detection and multi-class classification.

\bibliographystyle{IEEEtran}

\bibliography{reference/ref}

\end{document}